\documentclass[twocolumn,english,aps,pra,superscriptaddress,letterpaper]{revtex4-1}
\pdfoutput=1
\usepackage{amsmath,amssymb,graphicx}
\usepackage{textcomp} 
\renewcommand{\vec}[1]{\boldsymbol{#1}} 

\begin{document}
\title{Brillouin Scattering Self-Cancellation}

\author{O. Florez}
\author{P.F. Jarschel}
\author{Y.A.V. Espinel}
\author{C.M.B. Cordeiro}
\author{T.P. Mayer Alegre}
\author{G.S. Wiederhecker}
\author{P. Dainese}\email{dainese@ifi.unicamp.br}
\affiliation{Gleb Wataghin Physics Institute, University of Campinas, 13083-970 Campinas, SP, Brazil}

\date{\today}
\begin{abstract}
The interaction between light and acoustic phonons is strongly modified in sub-wavelength confinement, and has led to the demonstration and control of Brillouin scattering in photonic structures such as nano-scale optical waveguides and cavities. Besides the small optical mode volume, two physical mechanisms come into play simultaneously: a volume effect caused by the strain induced refractive index perturbation (known as photo-elasticity), and a surface effect caused by the shift of the optical boundaries due to mechanical vibrations. As a result proper material and structure engineering allows one to control each contribution individually. In this paper, we experimentally demonstrate the perfect cancellation of Brillouin scattering by engineering a silica nanowire with exactly opposing photo-elastic and moving-boundary effects. This demonstration provides clear experimental evidence that the interplay between the two mechanisms is a promising tool to precisely control the photon-phonon interaction, enhancing or suppressing it.
\end{abstract}
\maketitle 
\newcommand{\sinfo}[1]{(see supp. info. #1)}
\newcommand{\nocontentsline}[3]{}
\newcommand{\tocless}[2]{\bgroup\let\addcontentsline=\nocontentsline#1{#2}\egroup}
\newcommand{\ignore}[1]{}
\newcommand{\Gij}[3]{\frac{\left<\vec{\mathcal{E}}_{#1}|\delta\tilde #2|\vec{\mathcal{E}}_{#3}\right>}{\left<\vec{\mathcal{E}}_{#1}|\epsilon|\vec{\mathcal{E}}_{#3}\right>}}


\tocless{\section{Introduction}}
Brillouin scattering arises from the interaction between an electromagnetic wave and an acoustic wave. An incident photon at frequency $\omega $ is scattered into an up- or down-shifted photon with frequencies $\omega \pm {\Omega }$ due to the absorption or creation of a phonon with frequency ${\Omega }$, respectively. The ability to strongly confine both optical and acoustic modes in photonic structures has opened new opportunities to control their interaction~\cite{Kippenberg:2008jv,Dainese:2006kia}, and several new or enhanced applications have been explored, such as in photonic signal processing~\cite{Kippenberg:2008jv,Marpaung:2015hb,Li:2013bl,Kang:2011kk,Dainese:2006kia,CasasBedoya:2015il,Pagani:2014ih}, demonstration and tailoring of slow light devices~\cite{Herraez:2006in,Thevenaz:2008bo,Zhu:2007cn,Okawachi:2005cw}, actuation and control of photonic structures using optical forces~\cite{Wiederhecker:2009exa,VanThourhout:2010fc,Butsch:2012bxa}, observation of Raman-like scattering~\cite{Dainese:2006fh,Kang:2009ku}, on-chip Brillouin scattering~\cite{Merklein:2015kr,VanLaer:2015di,Shin:2013fra}, generation of frequency combs~\cite{Braje:2009eo,Butsch:2014jq} and new sensing applications~\cite{Beugnot:2011jk}. This interaction has also been a platform for fundamental physics experiments such as coherent phonon generation~\cite{Wiederhecker:2008gwa}, cooling mechanical modes to ground state~\cite{Chan:2011dya,SafaviNaeini:2012iha}, generation of squeezed light states~\cite{SafaviNaeini:2013cx}, and synchronization of micromechanical oscillators~\cite{Zhang:2012ksa}. 

\begin{figure}[h!]
\includegraphics[scale=0.95,bb=4.482000 3.456000 196.523994 324.518545]{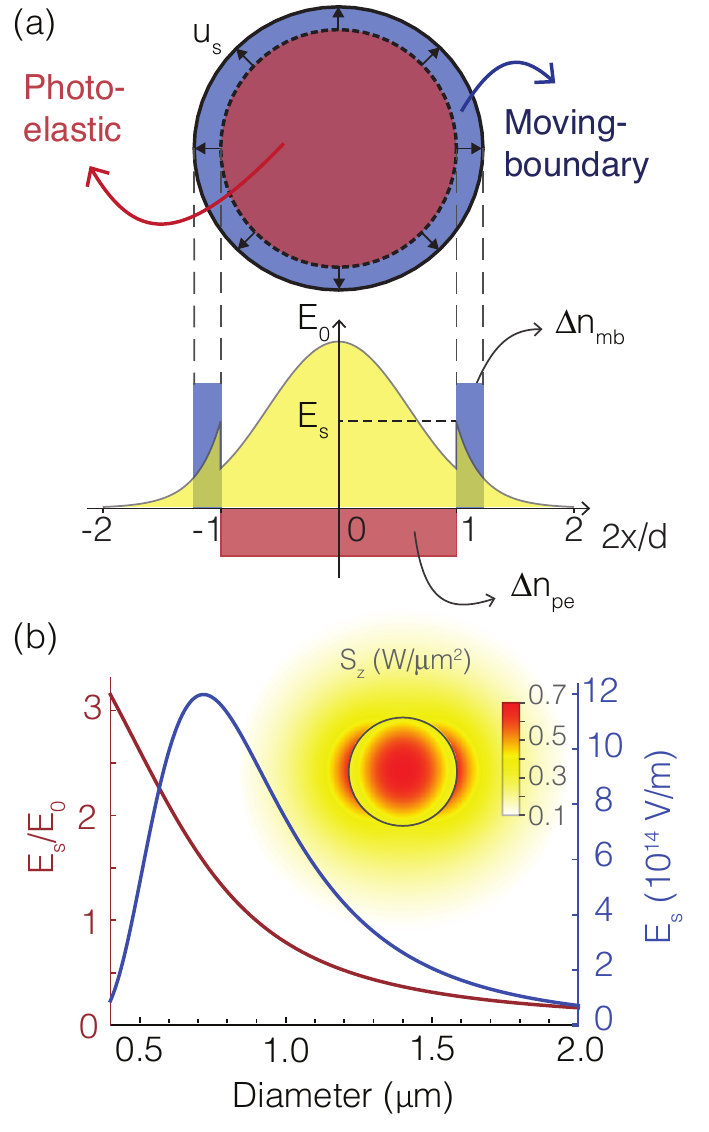}
\caption{\textbf{Illustration of the Photo-elastic and Moving-Boundary effects:} (a) refractive index perturbation due to the \textit{pe}-effect (red) and due to the \textit{mb}-effect (blue) in a nanowire under a linear radial expansion, overlaid with the electric field profile (\textit{d}: wire diameter, \textit{x}: horizontal coordinate, and \textit{u${}_{s}$}: surface displacement); (b) surface field amplitude \textit{E}${}_{s}$ as a function of wire diameter (blue: absolute value for 1~W optical power; red: amplitude relative to the center field \textit{E}${}_{0}$). Inset: Poynting vector for 0.55 \textmu m diameter.}	
\label{fig:fig1}
\end{figure}

Traditionally, the most common interaction mechanism between the optical and acoustic fields in low-contrast waveguides is the photo-elastic effect (\textit{pe}-effect), a well-known volume-effect in which the acoustic strain-fields perturb the material's refractive index~\cite{Boyd:2008tl}. For example, this effect dominates Brillouin scattering in standard optical fibers~\cite{Kobyakov:2010kp}. However, as the surface-to-volume ratio becomes larger and the index contrast higher, such as in nano-optical waveguides and cavities, the optical field experiences an additional surface effect due to the vibrating boundary~\cite{Rakich:2012eta,Johnson:2002jd,Rakich:2010hg,Wolff:2015jia}, referred to as moving-boundary effect (or \textit{mb-}effect). The schematic in Figure~\ref{fig:fig1}a illustrates the interplay between these two mechanisms in the case of a cylindrical-wire waveguide under a purely radial acoustic expansion.

We can simplistically characterize the strength of the \textit{mb}- and \textit{pe}-effects by evaluating their index perturbation \textit{depth}$\mkern1mu{\cdot}$\textit{area} products. As the wire boundary expands out by a small acoustic displacement $u_s$, the refractive index in a small region near the surface abruptly \textit{increases} by $\Delta n_{mb}=n_\text{glass}-n_\text{air}=0.45$ (where $n_{glass}$=1.45 and $n_\text{air}=1.0$ are the refractive indices of the silica wire and air cladding, respectively). Therefore the strength of the moving-boundary effect is: ${\eta }_{mb}={{\Delta }n}_{mb}{{A}}_{mb}=\left(n_\text{glass}-n_\text{air}\right)\pi du_s$, where ${{A}}_{mb}=\pi du_s$ is the area in which the \textit{mb}-effect effect takes place (here $d$ is the wire diameter). Although $\Delta n_{mb}$ is a relatively large perturbation, it affects only a small area (for thermally excited acoustic waves in the GHz frequency range, the boundary displacement is of the order of 10${}^{-16}$~m). Clearly, the \textit{mb}-effect is larger the higher the index contrast is. The photo-elastic effect, on the other hand, induces a perturbation throughout the entire wire cross-section, and so ${{A}}_{pe}={\pi d^2}/{4}$. Despite its tensorial nature, the order of magnitude of the \textit{pe}-effect can be estimated by considering only the radial strain component $S_{rr}$ (any other deformation in the longitudinal or azimuthal direction is ignored here, however a rigorous analysis is presented further in this article). Assuming the radial acoustic displacement increases linearly with \textit{r} as $u_r=u_s{2r}/{d}$ (with $r$ being the radius coordinate), the strain becomes simply $S_{rr}={\partial }_ru_r={2u_s}/{d}$. This then results in a index change of ${{\Delta }n}_{pe}=-n^3_\text{glass}p_{11}\frac{u_s}{d}$~\cite{Boyd:2008tl}. In contrast to the \textit{mb}-effect, a radial expansion leads to a \textit{reduction} in the refractive index (this is true since silica glass has a positive photo-elastic coefficient $p_{11}=0.121$~\cite{Yariv_Yeh_1983}). The strength of the \textit{pe}-effect is then ${\eta }_{pe}=-n^3_\text{glass}p_{11}\pi du_s$. With that, we estimate that the ratio $\eta_{pe}/\eta_{mb} = -n^3_\text{glass} p_{11}/4\Delta n_{mb}\approx -0.2$. This order of magnitude estimate indicates that both effects are indeed comparable in silica nanowires and, moreover, can have opposite signs.

Obviously, this simple analysis overlooks the crucial role of the optical fields. Not only the field strength of both incident and scattered waves must be non-negligible in the region where the index perturbation occurs, but also its spatial profile plays as a weighing function for each perturbation in a first-order spatial average. Figure~\ref{fig:fig1}b shows the evolution of the field on the surface as a function of wire diameter for an optical wavelength of 1.55~\textmu m (the inset shows the intensity profile for a 0.55 \textmu m diameter, clearly distributed over an area larger than the wire itself). The surface field increases as the diameter is reduced, reaches a maximum and then decreases for very small diameter. This defines a region where the \textit{mb}-effect might become relevant in the photon-phonon interaction, somewhere between 0.5 and 2.0~\textmu m in diameter. This behavior allows us to scan the nanowire diameter until we find a point in which the optical field provides just the right weighing, so that the elasto-optic and the moving-boundary effects cancel out exactly. 

\begin{figure*}
\includegraphics[scale=1,bb=1.174781 7.488000 507.545985 280.997991]{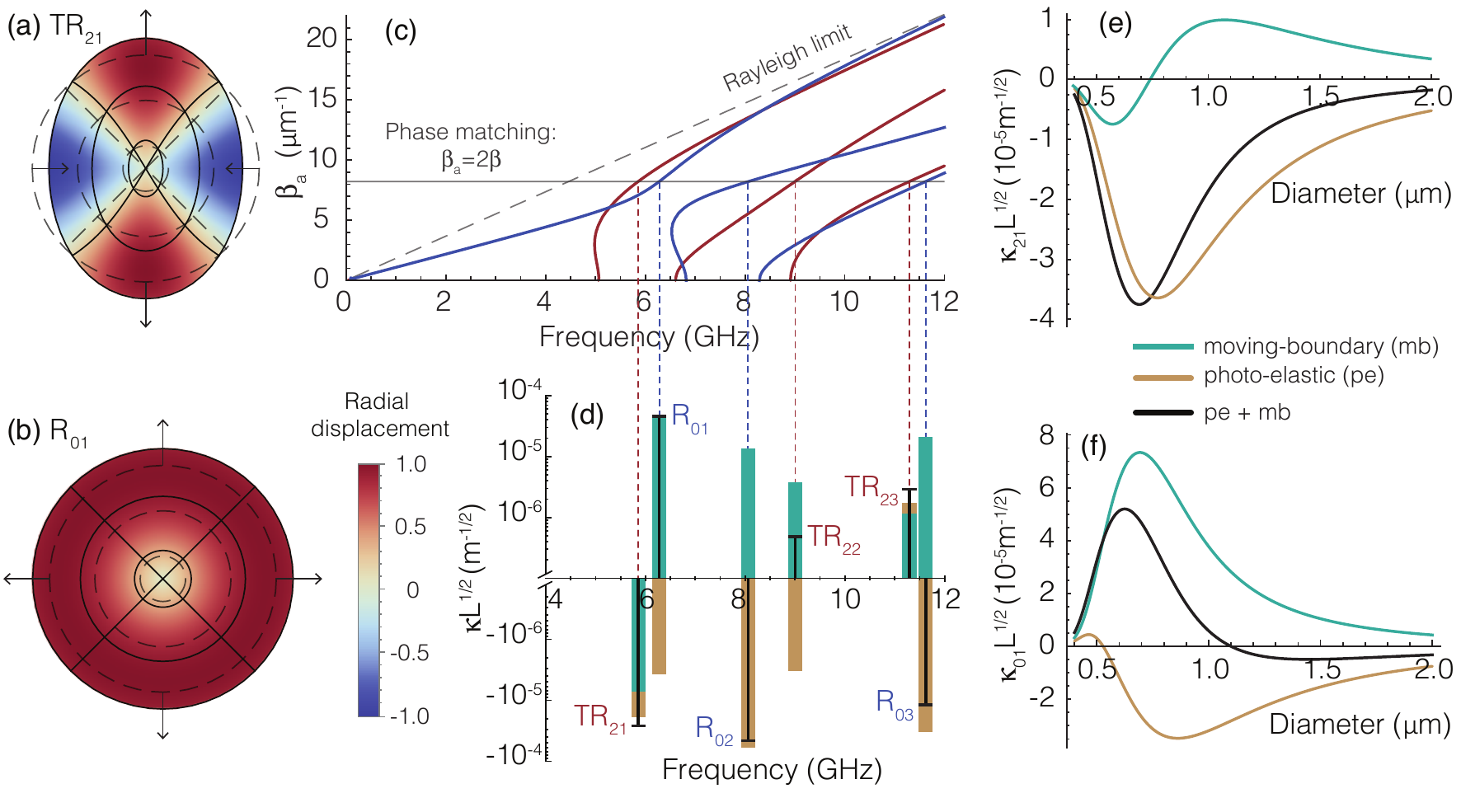}
\caption{\textbf{Acoustic modes and Perturbation strength:} nanowire cross-section under deformation due to the two fundamental\textit{ }Rayleigh acoustic modes: axially asymmetric torsional-radial \textit{TR}${}_{21}$ mode in (a) and axially symmetric radial \textit{R}${}_{0}$${}_{1}$ mode in (b). Solid and dashed lines represent respectively the deformed and un-deformed parametric lines, and the color profile represents the normalized radial acoustic displacement amplitude; (c) dispersion relation for both families (\textit{R}${}_{0}$${}_{m}$ and \textit{TR}${}_{2}$${}_{m}$) for a nanowire with 0.55 \textmu m diameter. The horizontal continuous line represents the phase-matching condition for backward Brillouin scattering; (d) coupling coefficients calculated for each phase-matched acoustic mode also for a nanowire with 0.55 \textmu m diameter(green: moving-boundary contribution; orange: photo-elastic contribution; and black: total \textit{pe} + \textit{mb}); (e) and (f) coupling coefficients as a function of the nanowire diameter. The coupling coefficients were calculated for acoustic modes normalized to thermal energy (at 300~K), and the factor ${L}^{1/2}$ scales the normalization to any waveguide length.}  	
\label{fig:fig2}
\end{figure*}

With these two effects simultaneously perturbing the dielectric constant, precise engineering of the optical and acoustic modes in a certain structure can be used to control their interaction, i.e. enhancing or suppressing it. In~\cite{Rakich:2010hg}, an analogous argument in terms of optical forces (electrostriction and radiation pressure) has been theoretically explored for a variety of structures and materials, demonstrating the richness of such interplay between the two mechanisms. In this paper, we demonstrate experimentally that we can achieve exact cancellation of backward Brillouin scattering by simply varying the wire diameter. It is interesting to note that physically all conditions favor strong photon-phonon interaction: (i) phase-matching condition is satisfied, (ii) both optical and acoustic fields are strongly confined and highly overlapping spatially and, most importantly, (iii) individually each interaction mechanism is strong. However, due to a precise control of the acoustic and optical mode profiles, the surface vibrations produces a dielectric perturbation that cancels out exactly the perturbation caused by internal body strain vibrations. We refer to this effect as Brillouin scattering \textit{Self}-Cancellation (BSC), since the cancellation arises from the same acoustic mode that creates each effect individually. Experimentally demonstrating the BSC-effect validates our fundamental understanding of the photon-phonon interaction in sub-wavelength confinement regimes, opening up the possibility to selectively suppress or further enhance the interaction by exploring both effects simultaneously.

The exact contribution from each mechanism can be calculated using standard perturbation theory (see Supplementary material for details). The strength of the photon-phonon coupling through photo-elastic effect is expressed mathematically as~\cite{Yariv_Yeh_1983}:
\begin{equation}
{\kappa }_{pe}=\frac{\omega{\varepsilon }_0}{8}\int{\vec{E}^*_s \mkern1mu{\cdot} {{\Delta }\vec{\epsilon} }^*_{pe} \mkern1mu{\cdot} \vec{E}_pdA},
\label{eq:kappa_pe}	
\end{equation}
where ${{\Delta }\vec{\epsilon} }_{pe}=-n^4 \vec{p}\mkern1mu{\cdot} \vec{S}$ is the relative permittivity perturbation caused by the acoustic strain tensor ($\vec{S}$), $n$ and $\vec{p}$ are the material's refractive index and photo-elastic tensor, while $\vec{E}_p$ and $\vec{E}_s$ are the power-normalized electric field profiles for the pump and scattered waves. Physically, the overlap integral simply represents a spatial average of dielectric perturbation weighted by the optical fields' profiles. As discussed before, the same acoustic wave that causes the \textit{pe}-effect, also causes a displacement on the boundaries that define the waveguide structure (in our case the wire boundary), and again using first order perturbation, the moving-boundary coupling coefficient is given by~\cite{Johnson:2002jd}:
\begin{equation}
{\kappa }_{mb} = \frac{\omega\epsilon_0}{8} \oint u_{\bot}(\Delta\epsilon_{mb} \vec{E}_{s,\parallel}^*\mkern1mu{\cdot}\vec{E}_{p,\parallel} +\Delta\epsilon_{mb}^{-1} \vec{E}_{s,\bot}^*\mkern1mu{\cdot}\vec{E}_{p,\bot})dl,
\label{eq:kappa_mb}
\end{equation}
where the perturbation coefficients are ${{\Delta }\epsilon }_{mb}=n^2_\text{glass}-n^2_\text{air}$ and $\Delta\epsilon_{mb}^{-1}=(n^{-2}_\text{air}-n^{-2}_\text{glass})n^4_\text{air}$, and $u_{\bot}$ is the acoustic displacement normal to the wire surface. In Equation~\ref{eq:kappa_mb}, the normal component of the electric field is evaluated in the outer region (air cladding) to correctly take into account the field discontinuity (this is equivalent to the definition in~\cite{Johnson:2002jd} using the continuous normal displacement field). Physically this line integral represents an average of the dielectric perturbation in an infinitesimal area along the waveguide perimeter, weighted by the optical fields. To first-order, the overall interaction strength is determined by sum of the coupling coefficients ${\kappa }={\kappa }_{pe}+{\kappa }_{mb}$ and the effect of Brillouin Self-Cancellation is achieved when ${\kappa }_{mb}={-{\kappa }}_{pe}$. 

In order to evaluate the coupling coefficients in Equations~\ref{eq:kappa_pe} and~\ref{eq:kappa_mb}, one must obtain the acoustic mode profiles. Each mode creates a different perturbation profile, both on the surface and throughout the wire cross-section. The acoustic modes in a cylindrical rod geometry can be categorized in symmetry-based modal families~\cite{Waldron:1969bb}, and can be calculated analytically. The fundamental optical mode interacts most efficiently with two acoustic mode families: axially symmetric radial (\textit{R}${}_{0}$${}_{m}$) modes and axially asymmetric torsional-radial (\textit{TR${}_{2}$${}_{m}$}) modes~\cite{Shelby:1985eb}. Figures~\ref{fig:fig2}a and~\ref{fig:fig2}b show the wire cross-section under the deformation caused by the fundamental modes, \textit{TR}${}_{2}$\textit{${}_{1}$} and \textit{R${}_{0}$${}_{1}$}, respectively. The acoustic dispersion relation calculated for a silica wire with 0.55 \textmu m diameter is shown in Figure~\ref{fig:fig2}c. The horizontal line represents the phase matching condition for backward Brillouin scattering ${\beta }_a=2\beta $ (where ${\beta }_a$ and $\beta$ are the acoustic and optical propagation constants, respectively), and the crossing points with the dispersion curves determine the frequencies of the acoustic modes involved in the interaction. Due to such small wire diameter (from 0.5 and 2.0 \textmu m), there are only a few acoustic modes per family, which leads to multi-peaked Brillouin spectrum~\cite{Dainese:2006kia}. 

\begin{figure}
\includegraphics[scale=1, bb=2.538000 1.422000 226.115993 300.005991]{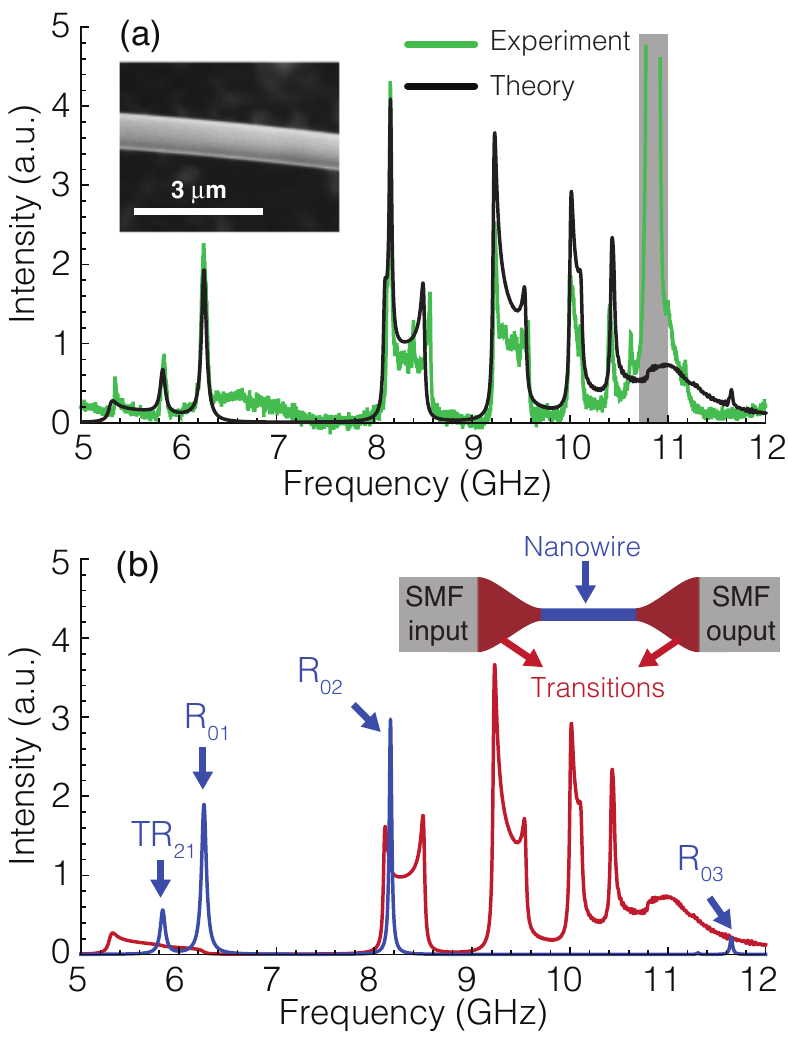}
\caption{\textbf{The Brillouin spectrum:} (a) experimental (green) and theoretical (black) Brillouin backscattering spectra for a sample with a diameter of 0.55 \textmu m. The shaded region represents the scattering peak due to the single-mode fiber (SMF) pigtails; (b) contributions to the total spectrum arising from the transition regions (red) and from the actual center nanowire (blue). The theoretical spectrum in (a) is the sum of these contributions. Insets: scanning electron microscope image of a nanowire and schematic of the sample structure, indicating the SMF input/output pigtails, the transition regions and the center nanowire.}	
\label{fig:fig3}
\end{figure}

The fundamental modes in each families, \textit{R}${}_{01}$ and \textit{TR}${}_{21}$, are of particular interest because they have the largest acoustic displacement near the surface (thus enhancing \textit{mb}-effect). In the high frequency limit -- when the acoustic frequency is much larger than the mode cutoff frequency -- the phase velocity of these two modes (\textit{R}${}_{01}$ and \textit{TR}${}_{21}$) approaches a certain limit, so-called \textit{Rayleigh speed} (and for that these modes are referred to as Rayleigh modes)~\cite{Auld:1992ub,Thurston:1978eqa}. Brillouin scattering due to such Rayleigh modes has been observed recently~\cite{Beugnot:2014cda}. In this limit, the Rayleigh speed is lower than the bulk longitudinal or transverse speeds, which in turn means that the transverse wavevector becomes imaginary and the acoustic field profile decays exponentially from the surface inwards~\cite{Auld:1992ub}. In other words, in the limit of high frequency, these modes are pure surface waves. Even below the Rayleigh limit, a large surface displacement is expected, as can be seen from the profile in Figures~\ref{fig:fig2}a and \ref{fig:fig2}b. The calculated coupling coefficients for this 0.55 \textmu m wire diameter are shown in Figure~\ref{fig:fig2}d, in which the individual contribution from each effect is shown separately for modes up to 12 GHz. Clearly, the shifting-boundary effect can not be neglected and, in fact, it dominates for some modes. Specifically, Figures~\ref{fig:fig2}e and \ref{fig:fig2}f show the calculated \textit{mb-} and \textit{pe}-coupling coefficients as a function of the wire diameter for the two Rayleigh modes (\textit{R}${}_{01}$ and \textit{TR}${}_{21}$). The \textit{pe}-effect dominates the interaction for the \textit{TR}${}_{21}$ mode throughout the entire diameter range explored here. There is no diameter in which \textit{mb}-effect compensates the \textit{pe}-effect, and therefore we do not expect to observe self-cancellation effect for \textit{TR}${}_{21}$. This can be easily understood based on the azimuthal ${{\cos} 2\phi \ }$ dependence, which means that the wire surface is deformed to a somewhat elliptical shape. The sign of the dielectric perturbation due to surface shift follows the radial displacement sign along the wire perimeter (i.e. positive in the regions the wire expands out and negative where it is contracted). As a result, the average \textit{mb}-perturbation (i.e. the line integral in $\kappa_{mb}$) is reduced. For the \textit{R}${}_{01}$ mode, on the other hand, the \textit{mb}- and \textit{pe}-effects are of the same magnitude and have opposite signs, as our simple analysis indicated. Clearly, at a diameter equals to 1.1 \textmu m, they exactly cancel out and no backward Brillouin scattering should be observed for this mode.
\tocless{\section{Experimental results} }
Samples of silica nanowires were fabricated by heating and stretching standard telecommunications fiber~\cite{Birks:1992ksa}. A geometric schematic is shown in the inset of Figure~\ref{fig:fig3}b. Two transition regions connect the actual nanowire (central region) to single-mode fibers at both input and output ends. All nanowires are approximately 8~cm long and have transition regions of approximately 5~cm on each side. We report results on samples with diameter ranging from $\sim$0.51 to 1.34~\textmu m. The precise wire diameter was determined using a non-destructive experimental technique based on forward Brillouin scattering. Both experimental setups for forward and backward Brillouin scattering characterization are described\ignore{briefly in Methods, and additional details are given} in the Supplementary materials.

The spectrum obtained for a sample with 0.55~\textmu m diameter is shown in Figure~\ref{fig:fig3}a, along with the theoretical spectrum calculated for the same nanowire diameter (the theoretical linewidth were obtained directly from the experimental spectrum). The first point to highlight is that several peaks in the experimental spectrum do not correspond to the frequencies of acoustic modes calculated for a wire with the same diameter (which were shown in Figure~\ref{fig:fig2}c). These measured additional peaks can be explained by considering the single-mode fiber pigtail and the transition regions. The single-mode fiber at the input gives origin to the peak at 10.8-10.9~GHz (shaded in gray as this is not the focus of this paper). In Figure~\ref{fig:fig3}b, we show the calculated contributions from the nanowire with uniform diameter (in blue) and from the transition region with varying diameter along its length\cite{Birks:1992ksa} (in red). The transition region leads to broad scattering bands, which arises because the phase-matching frequency sweeps the acoustic dispersion curve as the diameter varies. The theoretical curve in Figure~\ref{fig:fig3}a is the sum of these two curves from Figure~\ref{fig:fig3}b. The theoretical spectrum explains the one observed experimentally with remarkable agreement, and no fitting parameter was required (except for a 1\% adjustment on the bulk acoustic velocities to match the calculated and measured frequencies). 
\begin{figure*}
\includegraphics[scale=1,bb=0.774000 1.476000 509.669984 309.815991]{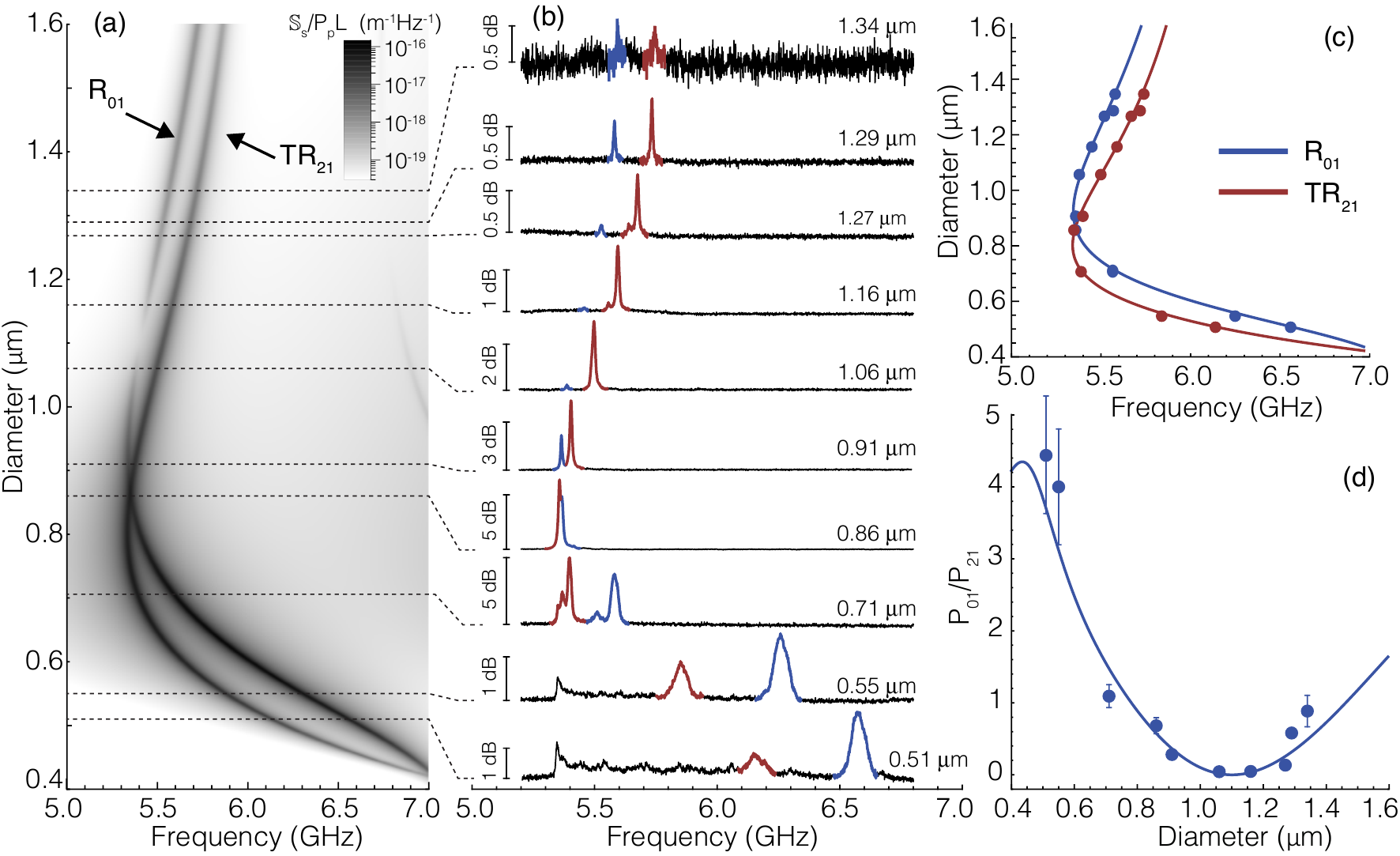}
\caption{\textbf{Observation of the Brillouin Self-Cancellation effect:} (a) theoretical evolution of the Brillouin backscattering spectrum as a function of the nanowire diameter for both Rayleigh modes \textit{R}${}_{01}$ and \textit{TR}${}_{21}$. The color-scale represents the scattered power spectral density (normalized by the pump power and wire length); (b) series of experimental spectra for nanowires with different diameters (as indicated in each spectrum). The peaks due to the \textit{R}${}_{01}$ or the \textit{TR}${}_{21}$ are colored in blue and red, respectively; (c) measured central frequencies for both \textit{R}${}_{01}$ and \textit{TR}${}_{21}$ (dotted blue and red points) compared with the expected theoretical frequencies (solid curves) as a function of diameter; (d) ratio between the \textit{R}${}_{01}$ and the \textit{TR}${}_{21}$ total scattered power for all samples, and the theoretical result (solid curve) given by $\kappa^2_{01}/\kappa^2_{21}$.}	
\label{fig:fig4}
\end{figure*}

To demonstrate Brillouin Self-Cancellation in our silica nanowires, we show in Figure~\ref{fig:fig4}a the calculated Brillouin spectrum evolution as a function of the nanowire diameter for the two fundamental acoustic modes (\textit{R}${}_{01}$ and \textit{TR}${}_{21}$). Slightly above their accidental frequency crossing (at \textit{d}~=~0.9~\textmu m), the BSC effect of the \textit{R}${}_{01}$ mode is predicted for a roughly 50~nm diameter range around 1.1~\textmu m. This behavior is precisely confirmed in the series of measured spectra, shown in Figure~\ref{fig:fig4}b. Each spectrum was obtained for a nanowire with a distinct diameter, as indicated. Each spectrum generally shows two well-defined peaks (except for the smallest diameters that more clearly shows a spectral band due to the transition regions). The two peaks correspond to the Rayleigh modes \textit{R}${}_{01}$ and \textit{TR}${}_{21}$ (colored in blue and red, respectively). Their central frequencies were measured and compared with the theoretically prediction in Figure~\ref{fig:fig4}c, with quite good agreement. The particular frequency versus diameter evolution is discussed in detail in the Supplementary material. The \textit{TR}${}_{21}$ peak is observed in all experimental spectra, in agreement with the theory, since there is no BSC expected for this mode. In contrast, the \textit{R}${}_{01}$ is observed for small and large diameters however it is not observed (within the experimental noise limit) in the region around 1.1~\textmu m -- a clear evidence of the Brillouin Self-Cancellation effect. This is more readily observed in Figure~\ref{fig:fig4}d, where the total scattered power in the \textit{R}${}_{01}$ mode (\textit{P}${}_{01}$) is normalized to the total scattered power in the \textit{TR}${}_{21}$ mode (\textit{P}${}_{21}$). Such normalization eliminates the uncertainty with respect to parameters that directly influence the absolute scattered power such as slightly differences in wire length, differences in the transition region optical attenuation, exact power at the wire input and the wire attenuation itself. The solid curve is the theoretical result, which is simply given by $\kappa^2_{01}/\kappa^2_{21}$. A remarkable agreement is observed and clearly the \textit{R}${}_{01}$ mode intensity approaches the noise limit at 1.1~\textmu m, confirming the BSC effect. 
\tocless{\section{Discussion}}

\begin{figure}
\includegraphics[scale=1,bb=6.300000 1.908000 221.669993 311.417990]{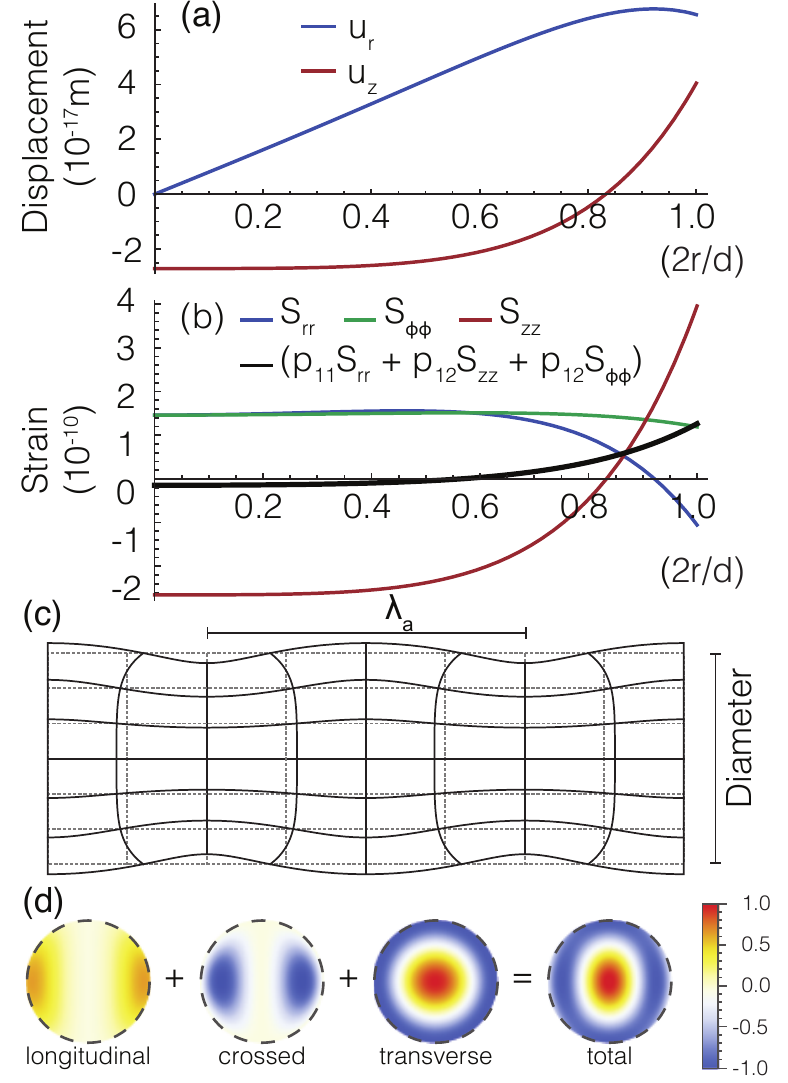}
\caption{\textbf{Physical understanding of the photo-elastic contribution:} (a) radial and axial displacement field profile for the fundamental \textit{R}${}_{01}$ acoustic mode for a wire with \textit{d} = 1.1 \textmu m diameter (absolute amplitude given by thermal energy normalization); (b) resulting strain profiles along with the total strain profile weighted by the photo-elastic coefficients; (c) illustrative wire deformation ($\lambda_{a}$ is the acoustic wavelength); (d) each integrand term in the coupling coefficient overlap integral (Equation~\ref{eq:kappa_pe}) is shown separately. The longitudinal term ($-{{\Delta }\epsilon }_{zz}{\left|E_z\right|}^2$) mostly cancels out the crossed term ($-2{{\Delta }\epsilon }_{rz}E_rE_z$.). As a result, the total perturbation profile is mostly dominated by the transverse term (${{\Delta }\epsilon }_{rr}{\left|E_r\right|}^2+{{\Delta }\epsilon }_{\phi \phi }{\left|E_{\phi }\right|}^2$).
 }	
\label{fig:fig5}
\end{figure}
To understand the cancellation effect in detail, it is essential to understand the acoustic profile and associated index perturbation profile. The fundamental \textit{R}${}_{01}$ mode has no dependence on the azimuthal angle $\phi $, and its displacement components are the axial $u_z{{\sin} \left({\beta }_{ac}z-{\Omega }t\right)\ }$ and the radial $u_r{{\cos} \left({\beta }_{ac}z-{\Omega }t\right)\ }$. Figure~\ref{fig:fig5}a shows the transverse profile for each component and Figure~\ref{fig:fig5}c illustrates the wire deformation under this particular mode (with amplitude arbitrarily large for better visualization). The maximum radial displacement, shown in Figure~\ref{fig:fig5}a, is near the wire surface and therefore induces significant moving-boundary effect. As mentioned before, the BSC effect shall occur only if the photo-elastic effect causes a negative net dielectric perturbation in order to compensate the positive \textit{mb}-effect perturbation. For the \textit{R}${}_{01}$ mode, the only nonzero strain components are the diagonal terms $S_{rr}={\partial }_ru_r$ (strain in the radial direction due to a radial displacement), $S_{\phi \phi }={u_r}/{r}$ (strain in the azimuthal direction due to a radial displacement) and $S_{zz}={\partial }_zu_z$ (strain in the axial-direction due to axial displacement), and the cross-term $S_{rz}={{\frac{1}{2}}}\left({\partial }_zu_r+{\partial }_ru_z\right)$. The strain profiles are shown in Figure~\ref{fig:fig5}b. The only nonzero terms in the dielectric perturbation tensor are also the diagonal terms ${{\Delta }\epsilon }_{rr}$, ${{\Delta }\epsilon }_{\phi \phi }$ and ${{\Delta }\epsilon }_{zz}$, and the cross-term ${{\Delta }\epsilon }_{rz}$. Explicitly, the diagonal term is ${{\Delta }\epsilon }_{rr}=-n^4\left(p_{11}S_{rr}+p_{12}S_{\phi \phi }+p_{12}S_{zz}\right)$. Similar expressions hold for the ${{\Delta }\epsilon }_{\phi \phi }$, ${{\Delta }\epsilon }_{zz}$ and ${{\Delta }\epsilon }_{rz}$ components. Note that the sign of the photo-elastic coefficients determines whether a positive (tensile) strain increases or reduces the dielectric constant. In silica (which is an isotropic medium), there are only two independent photo-elastic coefficients, \textit{p}${}_{11}$ and \textit{p}${}_{12}$, and they are both positives (except for the small crossed term ${{\Delta }\epsilon }_{rz}$ that depends on $p_{44}={{\frac{1}{2}}}\left(p_{11}-p_{12}\right)$, which is negative). Therefore an expansion leads to a reduction in the dielectric constant. More precisely, this means that ${{\Delta }\epsilon }_{rr}$ is negative if the sum of the strains (weighted by the photo-elastic coefficients) is positive. By examining the wire deformation profile in Figure~\ref{fig:fig5}c, it becomes very clear which regions have positive or negative strains fields. In the center of the wire, while it is compressed in the axial direction, it expands out radially (and thus azimuthally). This means that these strains fields counteract each other in the central region, leading to relatively smaller net contribution to the dielectric perturbation in the center of the wire, as shown in the black solid curve in Figure~\ref{fig:fig5}b. Note that this is just completely counter-intuitive for those accustomed to photo-elastic effect in conventional fibers, where axial strain completely dominates the effect. Near the wire surface, the strain competition picture changes radically. There is mostly simultaneous expansion in all (axial, azimuthal and radial) directions, and thus the strain fields contribute to a reduction on the dielectric constant (as required to achieve the BSC effect). One particular aspect for backward scattering (as opposed to forward scattering) is that $E_z$ -- the longitudinal component of the electric field -- changes sign for the backward wave. In cylindrical coordinates, the integrand in Equation~\ref{eq:kappa_pe} then becomes ${{\Delta }\epsilon }_{rr}{\left|E_r\right|}^2+{{\Delta }\epsilon }_{\phi \phi }{\left|E_{\phi }\right|}^2-{{\Delta }\epsilon }_{zz}{\left|E_z\right|}^2-2{{\Delta }\epsilon }_{rz}E_rE_z$. These terms are shown in the two-dimensional profile in Figure~\ref{fig:fig5}d (normalized to unit). For this particular diameter, the last two terms (longitudinal and crossed-terms) mostly cancel out each other, and the net result is dominated by the transverse terms ${{\Delta }\epsilon }_{rr}{\left|E_r\right|}^2+{{\Delta }\epsilon }_{\phi \phi }{\left|E_{\phi }\right|}^2$. Moreover, the transverse optical field profile provides the right balance between the positive (central) and negative (edges) perturbation regions. As a result, for a nanowire with 1.1~\textmu m diameter, the negative outer region dominates over the positive center region by just the exact amount necessary to cancel the positive dielectric perturbation due to the \textit{mb-}effect. This same dynamics explains why the \textit{pe}-effect has a zero-point near 0.51~\textmu m diameter (as previously shown in Figure~\ref{fig:fig2}f), when the positive central region just balances out the negative outer region. Note that Figure~\ref{fig:fig4}b shows the measured spectrum of a 0.51~\textmu m diameter nanowire, the point where the Brillouin scattering is \textit{totally} due to the moving-boundary effect. 

In conclusion, we have demonstrated experimentally the Brillouin Scattering Self-Cancellation effect in silica nanowires. Precise control of both optical and acoustic field profile turns out to create exactly opposing contributions due to photo-elastic and moving-boundary effects. In cylindrical geometry, we have demonstrated this effect for the fundamental axial-radial Rayleigh acoustic mode. A detailed understanding of the observed Brillouin spectrum and of the physical mechanism behind the cancellation effect is presented. We have also demonstrated that for 0.51 \textmu m diameter, the photo-elastic effect is zero (to first order) and the observed Brillouin scattering is produced \textit{completely} by moving-boundary effect only. Such rich interplay between photo-elastic and moving boundary effects can be further explored by modal engineering and the use of materials with different photo-elastic coefficients and refractive indices to observe the effect of Brillouin scattering Self-Cancellation in a variety of structure. Moreover, it can be used as a powerful tool to selectively control the photon-phonon interaction in photonic waveguides and cavities.  
\\
\vspace{-2pt}

\textbf{Acknowledgements:} this work was supported by the S\~ao Paulo Research Foundation (FAPESP) under grants 2013/20180-3, 2012/17765-7,  2012/17610-3, and 08/57857-2, and by the National Council for Scientific and Technological Development (CNPq), grant 574017/2008-9. O. Florez and Y.A.V. Espinel wish to thank the Coordination for the Improvement of Higher Education Personnel (CAPES) for financial support. Scanning Electron Microscopy was performed at the Center for Semiconductor Components (University of Campinas).

\onecolumngrid
\setcounter{figure}{0}
\setcounter{table}{0}
\setcounter{equation}{0}
\setcounter{section}{0}
\renewcommand{\theequation}{S\arabic{equation}}
\renewcommand{\thesection}{S\arabic{section}}
\renewcommand{\thesubsection}{\Alph{subsection}}
\renewcommand{\thesubsubsection}{\roman{subsubsection}}
\renewcommand{\thefigure}{S\arabic{figure}}
\renewcommand{\thetable}{S\arabic{table}}

\bibliography{dainese_sbs}

\newpage
\tocless{ \section*{Supplementary Information} }

\section{\label{sup:nlo}Nonlinear polarization}
The macroscopic Maxwell equations in the time domain for a nonmagnetic medium with no free-charges are: 
\begin{equation}
\begin{split}
{\nabla }\mkern1mu{ \cdot} \vec{D}&=0,\\
{\nabla }\mkern1mu{\cdot} \vec{H}&=0,\\
{\nabla }\times \vec{E}&=-{\mu }_0{\partial }_t \vec{H},\\
{\nabla }\times \vec{H}&={\partial }_t \vec{D}.
\label{eq:MaxEq}
\end{split}
\end{equation}

Here $\vec{D}={\varepsilon }_0\epsilon \vec{E}+\vec{P}_{NL}$ is the electric displacement field, ${\varepsilon }_0$ is the vaccum permittivity, $\epsilon $ is the relative dielectric permittivity and ${\mu }_0$ is the vacuum permeability. The nonlinear polarization $\vec{P}_{NL}$ induced by an acoustic wave is written in the time domain as:
\begin{equation*}
\vec{P}_{NL}\left(t\right)={\varepsilon }_0\Delta\vec{ \epsilon} \left(t\right)\mkern1mu{\cdot} \vec{E}\left(t\right),	
\end{equation*}
where ${\Delta }\vec{\epsilon} $ is the relative permittivity perturbation caused by the acoustic wave. For a harmonic perturbation, the relative permittivity perturbation can be written as:

\begin{equation*}
{\Delta }\vec{\epsilon} \left(t\right)={{\frac{1}{2}}}{\Delta }\vec{\epsilon} \left(x,y\right)e^{-i\left({\Omega }t-{\beta }_az\right)}+c.c.,
\end{equation*}
where ${\Delta }\vec{\epsilon} \left(x,y\right)$, ${\Omega }\ $ and ${\beta }_a$ are the perturbation transverse profile, angular frequency and propagation constant respectively. Since our interest is to investigate backward Brillouin scattering, we write the total field as the sum of a pump signal at $\omega_p$ and a backward propagating signal at $\omega_s$. We restrict our development to the Stokes line so that $\omega_s=\omega_p-{\Omega }$ (generalization to the anti-Stokes line at $\omega_{as}=\omega_p+{\Omega }$ is straightforward). We use the sub-index \textit{p} to denote the pump field (propagating in the forward direction) and \textit{s} to denote the scattered field propagating in the backward direction. We assume both fields as pure harmonics:

\begin{equation*}
\begin{split}
E_p\left(\vec{r},t\right)&={{\frac{1}{2}}}a_p\left(z\right)E_p\left(x,y\right)e^{-i\left(\omega_pt-{\beta }_pz\right)}+c.c.\\
E_s\left(\vec{r},t\right)&={{\frac{1}{2}}}a_s\left(z\right)E_s\left(x,y\right)e^{-i\left(\omega_st+{\beta }_sz\right)}+c.c..	
\end{split}
\end{equation*}

Note that the power carried by each signal is simply $P_n=s_n\frac{1}{2}{\left|a_n\right|}^2{Re}\ \left[\int{\left(E_n\times H^*_n\right)\mkern1mu{\cdot} \vec{z}dA}\right]$. From now on, $E_{p,s}$ denotes the transverse field profile $E_{p,s}\left(x,y\right)$. The factor $s_p=+1$ for the pump (co-propagating) and $s_s=-1$ for the Stokes signal (back-propagating) are used so that the optical power is always positive. We can always normalize the eigenmodes so that $s_n\frac{1}{2}{Re}\ \left[\int{\left(E_n\times H^*_n\right)\mkern1mu{\cdot} \vec{z}dA}\right]=1$~W, and then at any position along the waveguide the power carried by the eigenmode $n$ is simply $P_n={\left|a_n\right|}^2\mkern1mu{\cdot} 1$~W. We calculate the polarization generated by the pump field (i.e. neglecting the nonlinear polarization caused by the Stokes and anti-Stokes fields since in our experiments they are much weaker than the pump), obtaining:

\begin{equation*}
P_{NL}\left(t\right)=\frac{{\varepsilon }_0{\Delta }\epsilon }{4}a_pE_p\left(x,y\right)e^{-i\left[\omega_{as}t-\left({\beta }_p+{\beta }_a\right)z\right]}+\frac{{\varepsilon }_0{{\Delta }\epsilon }^*}{4}a_pE_p\left(x,y\right)e^{-i\left[\omega_st-\left({\beta }_p-{\beta }_a\right)z\right]}+c.c.	.
\end{equation*}
The first and the second terms are respectively the source terms for the anti-Stokes and Stokes scattered signals. Since we neglected the nonlinear polarization caused by the Stokes and anti-Stokes, there is no source term at the pump frequency $\omega_p$ (neither at high-order scattering such as $\omega_p\pm 2{\Omega }$). This approximation implies a constant pump field amplitude along the waveguide, $a_p\left(z\right)=a_p$ (in other words, pump depletion is neglected). 

\subsection{Perturbation theory}

The nonlinear polarization is treated as a perturbation term in Maxwell equations. We apply standard perturbation theory to calculate the evolution of the Brillouin backscattered signal amplitude $a_s\left(z\right)$ along the waveguide. In the linear regime ($P_{NL}=0$), the macroscopic Maxwell equations for a harmonic field $E e^{i\omega t}$ (specifically the curl equations in ~\ref{eq:MaxEq}) can be written in terms of operators as:

\begin{equation}
A\left.|\psi \right\rangle =-i{\partial }_zB\left.|\psi \right\rangle,
\label{eq:OperatorGenEq}
\end{equation}
where $\left.|\psi \right\rangle =\left[ \begin{array}{c}
E \\ 
H \end{array}
\right]$, and the operators are defined as:
\begin{align*}
A&=\left[ \begin{array}{cc}
\omega{\varepsilon }_0\epsilon  & -i{{\nabla }}_t\times  \\ 
i{{\nabla }}_t\times  & \omega{\mu }_0 \end{array}
\right], \\
B&=\left[ \begin{array}{cc}
0 & {-}\vec{z}\times  \\ 
\vec{z}\times  & 0 \end{array}
\right].
\end{align*}

For a waveguide, the solution to Equation~\ref{eq:OperatorGenEq} is of the form:

\begin{equation}
\left.|{\psi }_n\right\rangle ={{\frac{1}{2}}}a_ne^{i{\beta }_nz}\left.|{\varphi }_n\right\rangle\text{, with } \left.|{\varphi }_n\right\rangle =\left[ \begin{array}{c}
E_{\vec{n}}\left(x,y\right) \\ 
H_{\vec{n}}\left(x,y\right) \end{array}
\right],\label{eq:modalsol}
\end{equation}

where $E_{\vec{n}}$\textbf{ }and $H_{\vec{n}}$ are the field profiles for a specific eigenmode, and ${\beta }_n$ is the propagation constant (eigenvalue) at frequency $\omega $. In the linear regime $a_n$ is a constant. Substituting Equation~\ref{eq:modalsol} into~\ref{eq:OperatorGenEq}, we obtain a generalized eigenvalue problem:

\begin{equation}
	A\left.|{\varphi }_n\right\rangle ={\beta }_nB\left.|{\varphi }_n\right\rangle, 
	\label{Eq:WGEigenEq}
\end{equation} 
which for the back-propagating Stokes line becomes $A\left.|{\varphi }_s\right\rangle =-{\beta }_sB\left.|{\varphi }_s\right\rangle $ (where the negative sign arises naturally from the $e^{-i{\beta }_sz}$ dependence). The nonlinear polarization can be formally introduced as a perturbation term $\Delta A$ added to the operator $A$. For instance, from all the terms in the expression for the nonlinear polarization, the term oscillating at the Stokes field ($e^{-i\omega_st}$) is: 
\begin{equation*}
P_{NL}\left(t\right)=\frac{{\varepsilon }_0{{\Delta }\epsilon }^*}{4}a_pE_p\left(x,y\right)e^{-i\left[\omega_st-\left({\beta }_p-{\beta }_a\right)z\right]},	
\end{equation*}

In the presence of this term, Maxwell equations for the field oscillating as $e^{-i\omega_st}$ can again be cast in operator form as:

\begin{equation}
a_s e^{-i{\beta }_s z}A\left.|{\varphi }_s\right\rangle + a_p e^{i\left({\beta }_p-{\beta }_a\right)z}\Delta A\left.|{\varphi }_p\right\rangle =-i\left({\partial }_za_s-i{\beta }_sa_s\right)e^{-i{\beta }_sz}B\left.|{\varphi }_s\right\rangle,
\label{Eq:Aux1}
\end{equation}
where the perturbation operator is:

\begin{equation*}
\Delta A=\left[ \begin{array}{cc}
\frac{\omega_s{\varepsilon }_0}{2}{{\Delta }\epsilon }^* & 0 \\ 
0 & 0 \end{array}
\right].
\end{equation*}

Using the generalized eigenvalue Equation~\ref{Eq:WGEigenEq} for the Stokes signal, the first term on the left-hand side of Equation~\ref{Eq:Aux1} cancels with the second term on the right-hand side, which leads to:

\begin{equation}
\partial_z a_s = -i\kappa a_p e^{i\Delta \beta z}\text{, with } \kappa =\frac{\left\langle \varphi_s |\Delta A | \varphi_p \right\rangle}{4}.
\label{Eq:Aux2}
\end{equation}

The phase mismatch is $\Delta \beta ={\beta }_p+{\beta }_s-{\beta }_a$ and $\kappa $ is the coupling coefficient. In the definition of the coupling coefficient $\kappa $, we used the mode normalization
\begin{equation*}
	\left\langle \varphi_s | B | \varphi_s \right\rangle =2 \Re \left[\int{\left(E_n\times H^*_n\right)\mkern1mu{\cdot} \hat{z}dA}\right]=4s_s\text{[W]}=-4\text{[W]},
\end{equation*}
since $s_s=-1$. Note that this normalization guarantees that the unit of $\kappa $ is m$^{-1}$. The solution to Equation~\ref{Eq:Aux2} is:
\begin{align*}
a_s\left(0\right)&=-i\kappa a_p\frac{{1-e}^{i\Delta \beta L}}{i\Delta \beta },\\
P_s &=P_p{\left|\kappa \right|}^2L^2\text{sinc}^2\left(\frac{\Delta \beta L}{2}\right).	
\end{align*}

We have assumed that $a_s\left(L\right)=0$, where $L$ is the waveguide length, and $P_p$ is the pump power. In particular, for the case of perfect phase-matching $\Delta \beta =0$, the total backscattered signal amplitude and power are given by:
\begin{align}
a_s & = -ia_p\kappa L,\\
P_s & = P_p{\left|\kappa \right|}^2L^2.
\label{Eq:Pscatkappa}
\end{align}

Although one might be drawn to conclude that the scattered power increases with the square of the waveguide length $L$, this is not correct. We show below that acoustic normalization (with $k_BT$ of energy per mode) results in a $L^{-1/2}$ dependence for the coupling coefficient, which combined with the $L^2$ factor in Equation~\ref{Eq:Pscatkappa}, yields a linear dependence of scattered power on the waveguide length. 

The calculation of the inner product $\left\langle {\varphi }_s | \Delta A | {\varphi }_p\right\rangle $ in the coupling coefficient (Equation~\ref{Eq:Aux2}) must be performed with care. For the photo-elastic effect, the perturbation ${\Delta }\epsilon $ caused by a mechanical strain is sufficiently small so that $\kappa $ is simply
\begin{equation}
{\kappa }_{pe}=\frac{\omega_s{\varepsilon }_0}{8}\int{\vec{E}^*_s \mkern1mu{\cdot} {{\Delta }\vec{\epsilon} }^*_{pe} \mkern1mu{\cdot} \vec{E}_pdA},
\label{Eq:kappa_pe}	
\end{equation}
where ${{\Delta }\vec{\epsilon} }_{pe}=-n^4\vec{p}\mkern1mu{\cdot} \vec{S}$. However, in the case of a moving boundary, ${\Delta }\vec{\epsilon} $ is simply $n^2_1-n^2_2$, the difference in the relative permittivity between region 1 (wire core) and region 2 (air cladding). This perturbation is not small even for infinitesimal acoustic displacement and the field discontinuity must be taken into consideration. In this case, the correct expression for the coupling coefficient is (``Perturbation theory for Maxwell's equations with shifting material boundaries,'' Phys Rev E, 65 (2002)): 

\begin{equation}
{\kappa }_{mb}=\frac{\omega_s{\varepsilon }_0}{8}\oint{\left(\vec{u}\mkern1mu{\cdot} \hat{n}\right)\left[{{\Delta }\epsilon }^*_{mb}\vec{E}^*_{s,\parallel } \mkern1mu{\cdot}\vec{E}_{p,\parallel }+{{\Delta }\left({\epsilon }^{-1}_{mb}\right)}^*\vec{E}^*_{s,\bot } \mkern1mu{\cdot} \vec{E}_{p,\bot }\right]dl}	,
\label{Eq:kappa_mb}
\end{equation}
where ${{\Delta }\epsilon }_{mb}=n^2_1-n^2_2$ and ${{\Delta }\left({\epsilon }^{-1}_{mb}\right)}=\left(n^{-2}_2-n^{-2}_1\right)n^4_2$ and the normal component of the electric field is evaluated in the region 2 (air cladding) to correctly take into account the field discontinuity. Note that $\vec{u}\mkern1mu{\cdot} \hat{n}$ is the normal component of the acoustic displacement, which in a cylindrical nanowire is simply the radial displacement component $u_r$. The integral is performed along the waveguide circular boundary. Equations~\ref{Eq:kappa_pe} and~\ref{Eq:kappa_mb} were used to calculate the coefficients in Figures~\ref{fig:fig2}d-f. In these figures, $\kappa$ was calculated for acoustic modes normalized to thermal energy (at 300~K), and the factor $L^{1/2}$ scales the normalization to any waveguide length (this is developed in detail in the next section). For example, Figure~\ref{fig:fig2}d gives the product $\kappa L^{1/2}$ for a wire with with 0.55~\textmu m in diameter and so, for the \textit{R}${}_{0}$${}_{1}$ acoustic mode, $\kappa L^{1/2} = 4.5 \mkern1mu{\cdot} 10^{-5}$~m$^{-1/2}$. Using $L$ = 0.08~m, we then obtain $\kappa = 1.6\mkern1mu{\cdot} 10^{-4}$~m$^{-1}$. The total scattered power can be calculated using Equation~\ref{Eq:Pscatkappa}, and for a 1~W input pump power, the Stokes scattered power is approximately 170~pW.

\subsection{Acoustic mode normalization and scattered light spectrum}
Each acoustic mode carries a $k_BT$ of energy. Since the time average kinetic and potential energies are equal, we simply normalize the acoustic mode by making its average kinetic energy as ${\mathcal{E}}_k=\frac{1}{2}k_BT$. We write the acoustic field for a given mode identified with a sub-index $l$ as:
\begin{equation*}
{\vec{u}}_l\left(x,y,z,t\right)={{\frac{1}{2}}}u_l{\vec{U}}_l\left(x,y\right)\ e^{i\left({{\Omega }}_lt-{\beta }_lz\right)}+c.c.,
\end{equation*}
where, $u_l$ is the field amplitude (units of m) and $U_l\left(x,y\right)$ is the transverse mode profile (adimensional) normalized so that $\max{\left|U_l\left(x,y\right)\right|}=1$, and ${\beta }_l$ is the propagation constant of mode $l$ evaluated at the acoustic frequency ${{\Omega }}_l$. The average kinetic energy is then:
\begin{equation*}
{\mathcal{E}}_k=\int{\frac{1}{2}\rho \left\langle {\left|\frac{\partial {\vec{u}}_l}{\partial t}\right|}^2\right\rangle dV}=\frac{1}{4}\rho {{\Omega }}^2_l{\left|u_l\right|}^2L\int{{\left|U_l\left(x,y\right)\right|}^2dA}\\
\Rightarrow {\left|u_l\right|}^2=\frac{4{\mathcal{E}}_k}{\rho {{\Omega }}^2_lA_lL}	,
\end{equation*}
where ${\mathcal{E}}_k=\frac{1}{2}k_BT$ and $A_l=\int{{\left|U_l\left(x,y\right)\right|}^2dA}$. The last integral is performed over the waveguide cross-section area. We redefine the coupling coefficient so that the dependence on the acoustic amplitude $u_l$ (determined by thermal energy) becomes explicit:
\begin{equation*}
	\kappa =k u_l.
\end{equation*}
In this way, $k$ is fully determined by the modal profiles (optical and acoustic), and has unit of m${}^{-2}$. Explicitly, for the moving boundary effect we have:
\begin{equation}
	k_{sb}=\frac{\omega_s{\varepsilon }_0}{8}\oint{dl\left({\vec{U}}_l\mkern1mu{\cdot} \hat{n}\right)\left[\left(n^2_1-n^2_2\right)\vec{E}^*_{s,\parallel }\mkern1mu{\cdot} \vec{E}_{p,\parallel }+\left(n^2_1-n^2_2\right)\frac{n^2_2}{n^2_1}\vec{E}^*_{s,\bot }\mkern1mu{\cdot} \vec{E}_{p,\bot }\right]dl}.
\end{equation}
Similarly, for the elasto-optic effect we have:
\begin{equation}
k_{eo}=\frac{\omega_s{\varepsilon }_0}{8}\int{\vec{E}^*_s\mkern1mu{\cdot} \left(-n^4\vec{p}\mkern1mu{\cdot} \vec{\mathcal{S}}\right)\mkern1mu{\cdot} \vec{E}_pdA},	
\end{equation}
where ${\mathcal{S}}_{ij}{=}\frac{1}{2}\left(\frac{\partial }{\partial x_j}U_{l,i}\ e^{-i{\beta }_lz}+\frac{\partial }{\partial x_i}U_{l,j}\ e^{-i{\beta }_lz}\right)$ is the strain created by the acoustic mode $l$  divided by the thermal amplitude of the field $u_l$. With this definition, the total scattered power (assuming perfect phase-matching) becomes:
\begin{equation}
P_s=P_p{\left|k_l\right|}^2L\frac{4{\mathcal{E}}_k}{\rho {{\Omega }}^2_lA_l}=P_p{\left|k_l\right|}^2L\frac{2k_BT}{\rho {{\Omega }}^2_lA_l}.
\label{Eq:PscattTot}
\end{equation}
In deriving Equation~\ref{Eq:PscattTot}, we assumed that the acoustic field is harmonic and therefore the spectrum of the scattered light is single-frequency. In the presence of dissipation, the acoustic energy will be distributed over a Lorentzian spectrum. In order to calculate backscattered light spectrum, we simply write the acoustic energy spectrum density as
\begin{equation*}
\mathbb{E}\left({\Omega }\right)={\mathcal{E}}_k\frac{{{\Gamma }}/{\pi }}{{\left({\Omega }-{{\Omega }}_l\right)}^2+{{\Gamma }}^2},
\end{equation*}
so that $\int^{\infty }_0{\mathbb{E}\left({\Omega }\right)d{\Omega }}={\mathcal{E}}_k$, and ${{\Gamma }}/{2\pi }$ is the Brillouin linewidth (in units of Hz). Therefore, the power spectrum density of scattered light becomes:
\[{\mathbb{S}}_{s,l}\left(\omega \right)=P_p{\left|k_l\right|}^2L\frac{2k_BT}{\rho {{\Omega }}^2_lA_l}\frac{{{\Gamma }}/{\pi }}{{\left(\omega -\omega_p+{{\Omega }}_l\right)}^2+{{\Gamma }}^2}.\] 
For an instrument with frequency resolution ${\Delta }f$ (in units of Hz), the measured power (in W) in a 1 Hz resolution will be:
\begin{equation*}
 P_s[\text{W in bandwidth }{\Delta }f{]}={\mathbb{S}}_{s,l}\left(\omega \right){\Delta }\omega {=}{\mathbb{S}}_{s,l}\left(\omega \right){2}\pi {\Delta }f=P_p{\left|k_l\right|}^2L\frac{2k_BT}{\rho {{\Omega }}^2_lA_l}\frac{{2}{\Gamma }{\Delta }f}{{\left(\omega -\omega_p+{{\Omega }}_l\right)}^2+{{\Gamma }}^2}.	
\end{equation*}
Finally, the scattered power per unit of frequency (W/Hz) normalized by the pump power and the waveguide length is:
\begin{equation*}
	\frac{{\mathbb{S}}_s}{P_pL}={\left|k_l\right|}^2\frac{2k_BT}{\rho {{\Omega }}^2_lA_l}\frac{{2}{\Gamma }}{{\left(\omega -\omega_p+{{\Omega }}_l\right)}^2+{{\Gamma }}^2}.	
\end{equation*}
This result is used to generate the theoretical Brillouin backscattering spectrum in Figures~\ref{fig:fig3}a,~\ref{fig:fig3}b and~\ref{fig:fig4}a.

\newpage
\section{\label{sup:exp_setup}Experimental setup} 

\subsection{Diameter characterization}

Forward Brillouin scattering arises from an acoustic wave that oscillates transversally at the cut-off point (${\beta }_a=0$). Therefore, the observed Brillouin frequency shift is exactly the acoustic cut-off frequency. For a cylindrical rod, the cut-off frequency is inversely proportional to the wire diameter~\cite{Shelby:1985eb,Kang:2008eo,Dainese:2006fh}, and can be obtained by applying the free-surface boundary condition. For example, for the axially asymmetric torsional-radial (\textit{TR${}_{2}$${}_{m}$}) mode, \textit{f}${}_{m}$ = 2\textit{y${}_{m}$v${}_{T}/\pi$d}, where \textit{y${}_{m}$} are the numerical solution of the transcendental equation:

\begin{equation*}
\begin{split}
& J_1(y) \left(8 \left(\alpha ^2-1\right) \left(y^2-6\right) J_1(y \alpha )+4 \alpha  y \left(y^2-6\right) J_0(y \alpha )\right)+\\
&J_0(y) \left(2 y \left(\alpha ^2 \left(y^2+12\right)+y^2-24\right) J_1(y \alpha )-\alpha  y^2 \left(y^2-24\right) J_0(y \alpha )\right)=0
\end{split}	
\end{equation*}

Note that this equation depends only on the parameter $\alpha$ = \textit{v${}_{T}$/v${}_{L}$}, the ration between the shear and longitudinal bulk acoustic velocities, which for silica is approximately 0.63. The numerical solution for the fundamental \textit{TR${}_{2}$${}_{1}$} mode is \textit{y${}_{1}=2.34$}. Therefore, by measuring the forward Brillouin scattering frequency shift for the \textit{TR${}_{2}$${}_{1}$} mode, we can accurately determine the wire diameter by \textit{d} = 2\textit{y${}_{1}$v${}_{T}/\pi$ f${}_{1}$}.

\begin{figure}[h!]
\includegraphics[scale=1,bb=28.277999 9.918000 460.782478 165.677479]{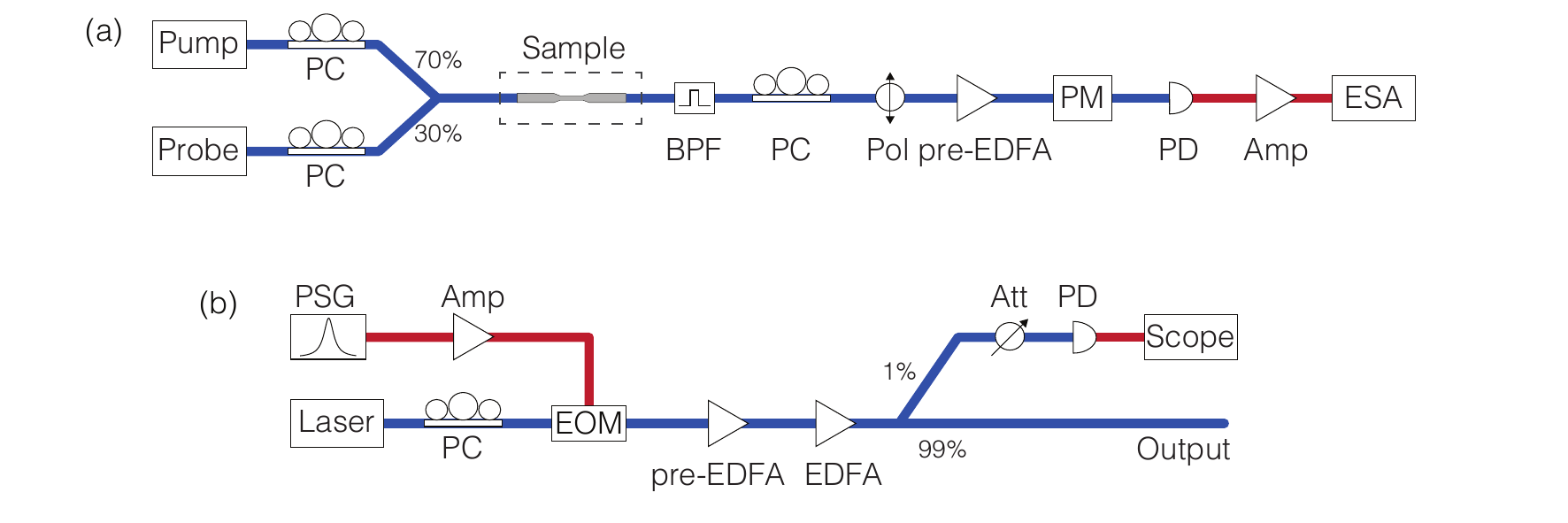}
\caption{Experimental setup used to characterize the forward Brillouin spectrum is shown in (a). The pulsed pump is generated using the setup in (b). PC: polarization controller; AMP: electrical amplifier; EOM: electro-optic modulator; EDFA (erbium-doped fiber amplifier); PM: in-line powermeter; BPF: band-pass optical filter; Pol: polarizer; Att: optical attenuator; PD: photodiode; and ESA: electrical spectrum analyzer.}	
\label{fig:figS2}
\end{figure}

The experimental setup used to characterize the forward Brillouin spectrum on our nanowires is based on a pump-and-probe technique, and is shown in Figure~\ref{fig:figS2}a. The fundamental \textit{TR}${}_{21}$ acoustic mode is excited using a pulsed laser source (pulse duration of 25 ps, repetition rate of 1 MHz and wavelength at 1570 nm). A probe continuous laser operating at 1550 nm is combined with the pump signal and launched into the silica nanowire. At the output, the pump laser is filtered out and the probe inserted into a polarizer in order to convert polarization modulation into amplitude modulation. The modulated signal is then amplified with an optical pre-amplifier and detected in a high-speed photodiode. Finally, the signal is analyzed in an electrical spectrum analyzer, and the peak frequency determined. The pulsed pump signal was generated using an Pulse Signal Generator and an Electro-Optical amplitude modulator. After modulation, the signal is pre-amplified using an pre-EDFA and then injected into a high-power EDFA (this is simply to saturate the high-power amplifier and reduce spontaneous emission). A small fraction $\sim$1\% of the pump power is detected using a photodiode and monitored in a oscilloscope.

\begin{figure}[h!]
\includegraphics[scale=1,bb=3.222000 1.098000 153.719995 125.021457]{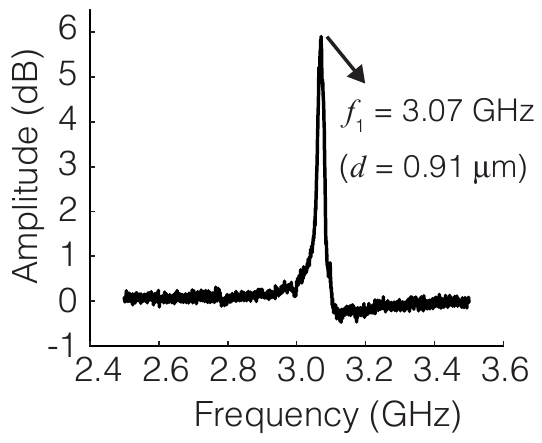}
\caption{Forward Brillouin spectrum arising from the fundamental \textit{TR}${}_{21}$ acoustic mode. From the measured peak frequency (3.06 GHz), the wire diameter was determined to be 0.91 \textmu m.}
\label{fig:figS3}
\end{figure}

Figure~\ref{fig:figS3} shows an example of the forward Brillouin spectrum measured. The peak observed is due to the fundamental \textit{TR}${}_{21}$ mode and the measured frequency is 3.07 GHz, from which the wire diameter was determined to be 0.91 \textmu m. All samples were characterized using this methods.

\newpage

\subsection{Brillouin backscattering experimental setup}

\begin{figure}[h!]
\includegraphics[scale=1,bb=1.850484 1.548000 231.887524 77.183998]{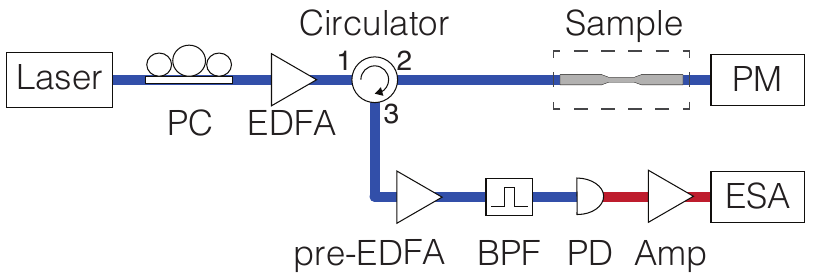}
\caption{Experimental setup used to characterize the Brillouin backscattering spectrum. PC: polarization controller; EDFA (erbium-doped fiber amplifier); PM: powermeter; BPF: band-pass optical filter; PD: photodiode; AMP: electrical amplifier; and ESA: electrical spectrum analyzer.}	
\label{fig:figS1}
\end{figure}

The setup used to characterize the backward Brillouin spectrum is similar to the one described in~\cite{Dainese:2006kia}, and is shown in Figure~\ref{fig:figS1}. A 1550~nm narrow-linewidth diode laser ($\sim$100~kHz linewidth) was amplified and launched into the silica nanowires using a circulator. The backscattered Brillouin signal is collected on port 3 of the circulator. Along with the frequency shifted Brillouin signal, a small linear reflection of the pump signal (non-frequency shifted) is used as a reference for heterodyne detection. These two signals (reference and Brillouin signal) are amplified using a low-noise Erbium doped fiber pre-amplifier, detected in a high-speed PIN photodiode ($>20$~GHz bandwidth), amplified electrically in a low noise radio-frequency pre-amplifier and dispersed in an electrical spectrum analyzer. 

\section{\label{sup:fem} Brillouin frequency shift \textit{vs}. wire diameter} 

\begin{figure}[h!]
\includegraphics[scale=1,bb=1.440000 0.414000 232.991993 251.531992]{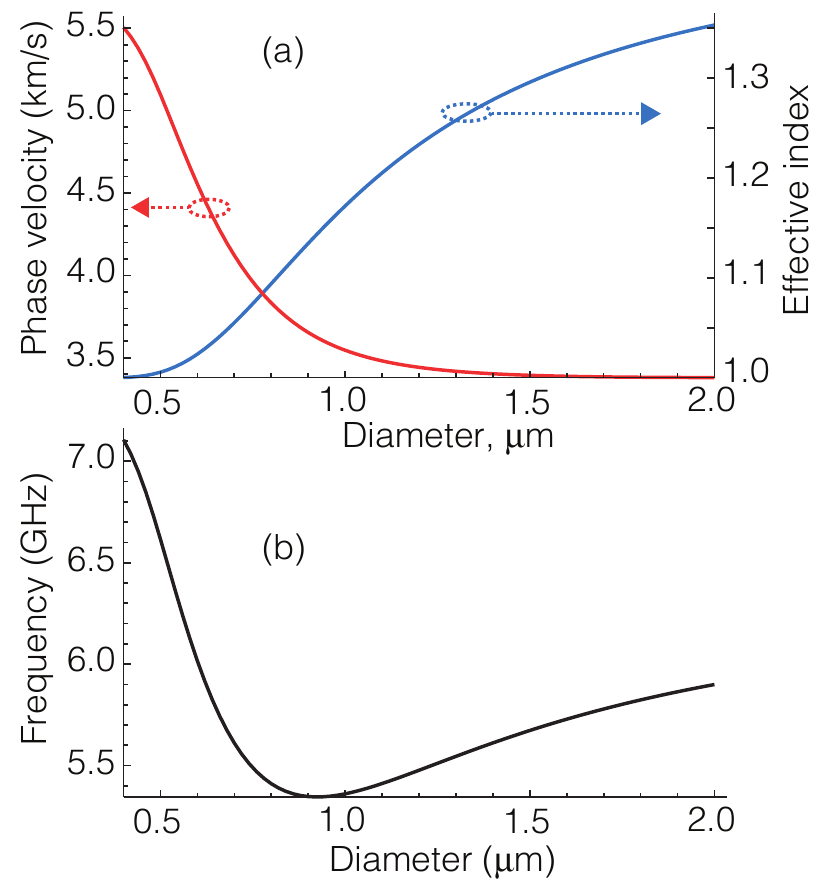}
\caption{ (a) Optical mode effective index and acoustic mode phase velocity for the \textit{R${}_{01}$} mode as a function of wire diameter. In (b), the resultant Brillouin frequency shift is shown as a function of diameter.}	
\label{fig:figS4}
\end{figure}

The evolution of the Brillouin frequency with diameter shown in Figure~\ref{fig:fig4}c can be understood by evaluating both optical and acoustic dispersion relations. The phase-matched acoustic frequency is \textit{f}${}_{ac}$ = 2\textit{n${}_\text{eff}$v${}_{ac}/\lambda$}, where $\lambda$ is the optical wavelength, \textit{n${}_\text{eff}$} is the optical mode effective index and \textit{v${}_{ac}$} is the phase velocity for a given acoustic mode. Therefore, the Brillouin frequency behavior depends on how \textit{n${}_\text{eff}$} and \textit{v${}_{a}$}${}_{c }$vary with the nanowire diameter, as shown in Figure~\ref{fig:figS4}a. The effective index decreases monotonically as the diameter is reduced due to increasing diffraction that spreads the optical energy out into the air cladding. Thus the effective index contributes to reduce the acoustic frequency \textit{f${}_{ac}$} as the diameter is reduced. On the other hand, the acoustic phase velocity \textit{v${}_{a}$}${}_{c}$ always increases as the diameter is reduced (from the lower limit referred to as Rayleigh velocity \textit{v${}_{R}$} to its maximum bulk longitudinal value \textit{v${}_{L}$}), and thus its contribution is to increase the acoustic frequency \textit{f${}_{ac}$}. The resultant frequency is shown in Figure~\ref{fig:figS4}b. Note that \textit{f}${}_{ac}$ first decreases before it increases again when the diameter is reduced (from right to left in Figure~\ref{fig:figS4}b). This behavior has its origin in the fact that that diffraction of the optical mode into the cladding occurs at a wire diameter comparable to the optical wavelength while the acoustic phase velocity is only affected when the wire diameter is comparable to the acoustic wavelength. Since the optical wavelength is twice the acoustic wavelength (as required by the phase-matching condition), it means that the reduction of \textit{n${}_\text{eff}$} impacts the Brillouin frequency \textit{f}${}_{ac}$ before \textit{v${}_{ac}$} does. This trade-off between how fast the optical effective index drops and how fast the phase velocity of acoustic Rayleigh waves increases as diameter becomes smaller determines the behavior of the acoustic frequency observed in Figures~\ref{fig:fig4}a and ~\ref{fig:fig4}c. 

\end{document}